\documentclass[twocolumn,showkeys]{revtex4}

\usepackage{amsmath}
\usepackage{amsfonts}
\usepackage{amssymb}
\usepackage{graphicx}
\usepackage[normalem]{ulem}
\usepackage{cancel}
\usepackage{float}
\usepackage{xcolor}
\usepackage{epstopdf}

\newcommand{\im}{\mathrm{i}}
\providecommand{\PT}[0]{\mathcal{PT}} %

\providecommand{\emr}[0]{\varepsilon_{1m}} %
\providecommand{\emi}[0]{\varepsilon_{2m}} %
\providecommand{\edr}[0]{\varepsilon_{1d}} %
\providecommand{\edi}[0]{\varepsilon_{2d}} %

\begin{document}

\title{
Phase transition in $\PT$ symmetric active plasmonic systems
}
\author{M. Mattheakis$^{1,2}$, T. Oikonomou$^{1, 3}$, M. I. Molina$^4$, G. P. Tsironis$^{1,3,5}$}
\vspace{0.5cm}
\affiliation{
$^1$Crete Center for Quantum Complexity and Nanotechnology,  Department of Physics, University of Crete, PO Box 2208, 71003 Heraklion, Greece\\
$^2$School of Engineering and  Applied Sciences, Harvard University, Cambridge, MA 02138, USA \\
$^3$Department of Physics, Nazarbayev University, 53 Kabanbay Batyr Ave., Astana 010000, Kazakhstan\\
$^4$Departamento de F\'isica, Facultad de Ciencias, Universidad de Chile, Santiago, Chile\\
$^5$National University of Science and Technology MISiS, Leninsky prosp. 4, Moscow, 119049, Russia
}

\date{\today}
\begin{abstract}
\noindent Surface plasmon polaritons (SPPs) are coherent
electromagnetic surface waves trapped on an insulator-conductor
interface. The SPPs decay exponentially along the propagation due
to conductor losses, restricting the  SPPs propagation length to
few microns. Gain materials can be used to counterbalance the
aforementioned losses. We provide an exact expression for the
gain, in terms of the optical properties of the interface, for
which the losses are eliminated. In addition, we show that systems
characterized by lossless SPP propagation are related to $\PT$
symmetric systems. Furthermore, we derive an analytical critical
value of the gain describing a phase transition between lossless
and prohibited SPPs propagation. The regime of the aforementioned
propagation can be directed by the optical properties of the
system under scrutiny.  {Finally, we perform COMSOL
simulations verifying the theoretical findings.}
\end{abstract}


\maketitle

\section{Introduction}
\label{sec:intro}

A  light-matter interaction called surface plasmon polaritons
(SPPs) has gained the scientists' interest due to its unique
properties, such as control of electromagnetic energy in
subwavelength scales \cite{SPPsubOptics, review2005,
kaxirasNano2014,huang2007}, high sensitivity in dielectric
properties \cite{kaxirasNano2014,LL1, LL2}, negative refraction
and hyperbolic wave front  \cite{negRef,selffocus}. SPPs have been
applied  in nanophotonics, imaging, optical holography, nano
antennas, biosensing,  integrated circuits and metamaterials
\cite{LL2,nanoAntennas,holography,biosense1,biosense2}. Important
progress has been made in plasmonics with two-dimensional
materials, such as graphene and black phosphorus, where  the
plasmonic properties can be tuned by using chemical doping or
applying external gate voltage
\cite{kaxirasNano2014,graphenePRB,grapheneNatPhot2012,basovNanoLet2011,basovNatLet2012,reviewIEEE2013,prl2014BF}.
Moreover, plasmonic lenses, waveguides and  meta-materials based
on graphene  have already been applied
\cite{kaxirasNano2014,teraPRB, enghetaSc2011}. Last but not least,
multi-layers structures have been created by stacking
two-dimensional crystals one on top of another providing
surprising electronic and optical features
\cite{prl2014BF,grapheneNatPhot2012,grapSheetAPL}.

Near  plasma frequency $\omega_p$,  the electrons on the
surface of  metals or semiconductors are free to move sustaining collective oscillations \cite{review2005,economou, gainOSA2004, maier,valag2009,valag2011}. The
coupling between light and electron  oscillations allows the
creation of Transverse Magnetic (TM) Electromagnetic (EM) surface
waves, namely surface plasmon polaritons (SSPs). From the
mathematical point of view, SPPs are surface waves  bounded along
the interface between two  materials with sign reversed dielectric
permittivities, i.e. a dielectric-conductor interface, and  their
EM field decays exponentially away from the interface (evanescent
waves) \cite{review2005,economou, gainOSA2004, maier,valag2009,valag2011}.

In this work, we focus on plasmonic waveguides formed by  a planar
interface which consists of two semi-infinite layers with reversed
sign permittivities, namely a dielectric and a metal. The dispersion relation  which characterizes the SPPs propagation can be determined by Maxwell Equations (ME) as \cite{review2005,gainOSA2004,huang2007,maier}
\begin{equation}
\label{eq:dispersion}
\beta=k_0~n=k_0~\sqrt{\frac{\varepsilon_d \varepsilon_m}{\varepsilon_d+\varepsilon_m}},
\end{equation}
where $k_0=\omega/c$ is the free space wave number of the incident
excitation light of angular frequency $\omega$, $n$ is the plasmon
effective  refractive index, $\varepsilon_d$ and $\varepsilon_m$
the permittivity of dielectric and metal, respectively, and $c$ is
the speed of light in vacuum.

SPPs decay  exponentially along the interface as well, due to the
metal losses. In mathematical language,  the metal losses are
described by a negative imaginary part in the permittivity
function of the metal, i.e. $\varepsilon_m=-\emr -\im \emi$, where
$\emr,\emi
>0$. Consequently, the SPPs wave number $\beta$ becomes complex,
viz. $\beta=\beta'+\im \beta''$, where the imaginary part accounts
for  losses of SPPs energy.  The imaginary part $\Im[\beta]$
of Eq. (\ref{eq:dispersion})   determines the characteristic
propagation length $L$, which shows the rate of change of the
energy attenuation of SPPs along the propagation axis
\cite{gainOSA2004, maier,huang2007}, that is
 \begin{equation}
 \label{eq:propL}
 L=\frac{1}{2 \Im[\beta]} .
 \end{equation}

Gain materials, rather than passive dielectrics, have been used to
reduce the losses in SPPs propagation. These active materials are
characterized by a complex permittivity function, i.e.
$\varepsilon_d=\edr+\im\edi$ with $\edr,~\edi>0$, where the
imaginary part accounts for gain, that is, the dielectrics give
energy to the system counterbalancing the metal losses
\cite{gainOSA2004,huang2007,avrPRB2004,leonNatPho2010,leonNatPhot2012}.
In addition, active dielectrics have been used for exploring $\PT$
symmetry in optical systems \cite{huang2014, lupu2011,lupu2013,ptPRA2014}
characterized by the condition that $n(-x)=n^*(x)$, where $n$ and
$n^*$ the refractive index and its complex conjugate,
respectively; $x$ denotes the spatial coordinate along the
interface. Metamaterials with $\PT$ symmetric effective refractive
index can be constructed by the combination of gain dielectrics
and loss metals \cite{lupu2011,lupu2013,ptPRA2014}. What makes
$\PT$ symmetric media interesting is that they allow control over
EM field by tuning the gain and loss of the materials.

It has been already demostrated in
\cite{gainOSA2004,huang2007,leonNatPho2010,leonNatPhot2012} that
for a certain value of gain, the losses in SPPs propagation may
vanish. Consequently, the SPPs propagation constant $\beta$ as well as the effective refractive index n become real and therefore the $\PT$ symmetry is satisfied, since
$n$  does  not exhibit any spatial dependence along the interface.
Furthermore, Eq. (\ref{eq:dispersion}) states that a $\PT$
symmetric $n$ leads to infinite propagation length, viz. lossless
SPPs propagation.

In the present work, we investigate theoretically and numerically the $\PT$ symmetry  in active plasmonic systems. In Section \ref{sec:mainTheory}, we provide an explicit expression of the gain, namely $\varepsilon_{PT}$, for which the losses in SPPs
propagation have been eliminated. In addition, we find a critical
value $\varepsilon_c$ of $\varepsilon_\PT$, where SPPs wave number
$\beta$ and the effective SPPs refractive index $n$ shift from
real to imaginary regime, subsequently the $\varepsilon_c$ is a $\PT$ symmetry breaking point.  It is remarkable that it is a steep
phase transition from lossless to prohibited SPP propagation,
which offers the opportunity  to control whether SPPs propagate or
not by tuning the optical properties of the interface.  In Section
\ref{sec:drude}, we apply the theoretical results derived in the
previous Section on interfaces comprised of active dielectrics and
Drude metals. In Section \ref{sec:simulations}, we proceed with numerical simulations by solving the full system of ME in the frequency domain by using the commercial multiphysics software COMSOL, and we show that  lossless SPPs propagation corresponding to $\PT$ symmetry can be achieved in the presence of gain dielectrics. Finally,  concluding remarks are offered in Section
\ref{sec:conclusion}.

\section{$\PT$ and critical gain}
\label{sec:mainTheory}

In this Section, we calculate the exact expression of the
dielectric permittivity   gain counterpart $\edi$, for which the
SPPs propagate without losses in the dielectric-metal interface.

Plugging the complex structure of  the dielectric and metal
permittivity into  Eq. (\ref{eq:dispersion}),   function $n$ can
be  written in the ordinary complex form as \cite{bookComplexSqrt}
\begin{equation}
\label{eq:nExpanded}
n=  \sqrt{\frac{\sqrt{x^2+y^2}+x}{2}} + \im \ \text{sgn}(y)\sqrt{\frac{\sqrt{x^2+y^2}-x}{2}}
\end{equation}
where $\text{sgn}(y)$ is the discontinuous signum function and
\begin{subequations}\label{eq:x,y}
\begin{eqnarray}
\label{eq:x}
x:=\frac{\varepsilon_{1d}\|\varepsilon_m\|^2-
\varepsilon_{1m}\|\varepsilon_d\|^2}{\|\varepsilon_d+\varepsilon_m\|^2}\\
\label{eq:y} y:=\frac{\varepsilon_{2d} \|\varepsilon_m\|^2-
\varepsilon_{2m} \|\varepsilon_d\|^2
}{\|\varepsilon_d+\varepsilon_m\|^2}
\end{eqnarray}
\end{subequations}
with $\|\varepsilon_*\|$ denoting the norm of the complex number $\varepsilon_*$.

Considering  the plasmon effective index $n$  in Eq.
(\ref{eq:nExpanded}) in the $(x,\,y)$--plane, we  observe that a
lossless SPP propagation, i.e, $\Im[\beta]=0$, is warranted when
the conditions $y=0$ and $x> 0$ are simultaneously satisfied. For
$y=0$ and  $x<0$, although the imaginary part in Eq.
\eqref{eq:nExpanded} vanishes due to the signum function, its real
part  becomes imaginary, i.e. $\beta=\mathrm{i}\,\sqrt{|x|}$, which does not correspond to propagation SPP modes.
Studying the permittivity dependence of $x$ and $y$ in Eq.
\eqref{eq:x,y} and solving the condition $y=0$ with respect to the
dielectric gain part $\varepsilon_{2d}$ for
$\varepsilon_{d}\neq-\varepsilon_m$, we obtain two exact
solutions, i.e., $\varepsilon_{2d}\rightarrow
\varepsilon_{2d}^{\pm}$ of the form
\begin{equation}
\label{eq:PTgain} \varepsilon_{2d}^{\pm} =
\frac{\|\varepsilon_m\|^2}{2\varepsilon_{2m}} \left( 1 \pm
\sqrt{1- \left( \frac{2 \varepsilon_{1d}\varepsilon_{2m}}
{\|\varepsilon_m\|^2} \right)^2 } \right)
\end{equation}
The result in Eq. \eqref{eq:PTgain} is in agreement with the one derived in Refs.  \cite{gainOSA2004,huang2007} following yet a different derivation path.
Invoking the physical argument of the SPP wave bound to the
dielectric-metal interface, we read that only
$\varepsilon_{2d}^{-}$  is of physical relevance,  since
$\varepsilon_{2d}^{+}$ leads to waves radiating in the transverse
towards the interface direction \cite{gainOSA2004}.
Taking into account the, by definition, positive real domain of
$\varepsilon_{2d}^{-}$ and the dependence of the latter on the
metal-dielectric components, we read that the following inequality
has to be satisfied
\begin{equation}
\label{eq:conditionForRealGain}
\|\varepsilon_m\|^2 >
2\,\varepsilon_{1d}\,\varepsilon_{2m}
\end{equation}
For $\varepsilon_d=-\varepsilon_m$ both $x$ and $y$ diverge
exhibiting asymptotically the same image, thus
\begin{equation} \label{eq:singularity}
\lim_{\varepsilon_{d}\rightarrow-\varepsilon_{m}}\Re[\beta]
=\lim_{\varepsilon_{d}\rightarrow-\varepsilon_{m}}\Im[\beta]
\;\rightarrow\;+\infty
\end{equation}
This in turn means that the former complex point does not belong
to the set domain of the lossless SPP propagation since $y\neq0$.

Solving, on the other hand, the equation $x=0$  with respect to
the dielectric gain $\varepsilon_{2d}$ for
$\varepsilon_d\neq-\varepsilon_m$, we may determine the critical
value $\varepsilon_{c}$ distinguishing the regimes of lossless and
prohibited SPP propagation, namely
\begin{equation}
\label{eq:criticalGain}
\varepsilon_c=\varepsilon_{1d}\sqrt{\frac{\|\varepsilon_m\|^2}{\emr\varepsilon_{1d}
}-1}
\end{equation}
Equating  Eqs. \eqref{eq:PTgain} and \eqref{eq:criticalGain},
i.e., $y=0=x$, we obtain the condition
$\varepsilon_{1d}\,\varepsilon_{2m}\,(\varepsilon_{1d}-\varepsilon_{1m})=0$
which reduces to $\varepsilon_{1d}=\varepsilon_{1m}$, since
$\varepsilon_{1d},\,\varepsilon_{2m}>0$. Replacing the former
value of $\varepsilon_{1d}$ in Eqs. \eqref{eq:PTgain} and
\eqref{eq:criticalGain} we obtain
$\varepsilon_{2d}^{-}\neq\varepsilon_c$ and
$\varepsilon_{2d}^{-}=\varepsilon_c=\varepsilon_{2m}$ for
$\varepsilon_{1m}<\varepsilon_{2m}$ and
$\varepsilon_{1m}>\varepsilon_{2m}$, respectively. The former case
is obviously a contradiction. The latter case corresponds to the
singularity point in Eq. \eqref{eq:singularity}, where $x,y\neq0$,
thus it is a contradiction as well.
In other words, $x$ and $y$ do not become zero simultaneously,
implying that the critical value $\varepsilon_c$ is not an element
of the domain set of the wave number $\beta(y=0)\equiv \beta_0$.
This is in agreement with the propagation length $L$ in Eq.
\eqref{eq:propL}. Indeed, when $y=0$ then $L$ tends to infinity,
which means that the SPP wave number must exhibit a nonzero
value.
An even more interesting point, unveiled from the $y=0=x$
analysis, is the estimation of the  $\beta_0$  behaviour when
approaching the critical point,
$\varepsilon_{2d}^{-}\rightarrow\varepsilon_c$ ($y=0$ and
$x\rightarrow0$), described by Eq. \eqref{eq:singularity} for
$\beta\rightarrow\beta_0$ with
$\varepsilon_{1m}>\varepsilon_{2m}$.
 {Physically, the former point corresponds to the wave
electrostatic character of zero phase velocity, known in
literature as the surface plasmon mode \cite{maier}.}
Including the discontinuity at the point $\varepsilon_c$, the
entire codomain of $\beta_0$ is described as follows
\begin{equation}\label{eq:ReBeta}
\beta_0=\begin{cases} ~\text{Real}, &  \quad x > 0  \\
~\text{Imaginary}, &   \quad x < 0\\
~\text{Complex Infinity}, &   \quad x \rightarrow 0
\end{cases}
\end{equation}

In the general case of $\beta$, where
 $y$ (excluding the point $\varepsilon_d=-\varepsilon_m$) may take nonzero values as
well, we observe the following.
For $\varepsilon_{2d}<\varepsilon_{2d}^{-}$, the signum function
in Eq. (\ref{eq:nExpanded}) is negative since then $y<0$
implying $\Im[\beta]<0\;\Rightarrow\;L<0$. This means that the
imaginary part of $\beta$ accounts for losses and the SPP
amplitude decreases along the propagation surface. Reversely, for
$\varepsilon_{2d}^{-}<\varepsilon_{2d}<\varepsilon_{2d}^{+}$
yielding $y>0\;\Rightarrow\;\text{sgn}(y)>0$, we have
$\Im[\beta]>0\;\Rightarrow\;L>0$. In this case the imaginary part
of $\beta$ accounts for gain and the SPPs amplitude increases along
the propagation surface. In the special case of
$y=0\;\Rightarrow\;\varepsilon_{2d}=\varepsilon_{2d}^{-}$ studied
above, the SPP amplitude is constant along the propagation
surface.
This behaviour of the signum function fully explains the results
observed in Ref. \cite{gainOSA2004} regarding the SPP amplitude.

An interesting feature  of the lossless SPP propagation case, i.e., for $\emi^{-} <\varepsilon_c$, in
regard to the refractive index $n$  is that the latter fulfils the
condition $n(y=0)=n^*(y=0)$, since its imaginary part vanishes
owing to the signum function. This in turn, may  be considered as
the $\PT$ symmetry phase condition, where $n$ is spatial
independent. However the structure is not $\PT$ symmetric in the narrow sense, the real value of the supported propagation constant along the interface admits time-reversal and geometrical symmetry. Then, the dielectric gain expression
$\varepsilon_{2d}^{-}$ in Eq.  \eqref{eq:PTgain} can be attributed
to the $\PT$ symmetry property satisfied by the lossless SPP propagation and
denoted as $\varepsilon_{2d}^{-}\equiv \varepsilon_{\PT}$. We
shall keep this denomination in what follows. On the contrary, in the case $\emi^{-} <\varepsilon_c$, the $\PT$ condition is not satisfied, since the refractive index is imaginary. Subsequently, the critical gain $\varepsilon_c$ may be regarded as the $\PT$-symmetry breaking point of the plasmonic system under scrutiny.

\section{Active dielectric - Drude metal interface}
\label{sec:drude}

It is quite remarkable that  $\varepsilon_\PT$ as well as
$\varepsilon_c$  depend on the optical  properties of the
dielectric and metal, that is, $\edr$, $\emr$ and $\emi$.
The metal permittivity in turn  may generally exhibit a dependence
on the angular frequency $\omega$, so that by tuning $\omega$  we
may control the values of $\varepsilon_\PT$ lying below or above
$\varepsilon_c$.
Precisely, for interfaces comprised of  an active dielectric and a
Drude metal \cite{maier}, $\varepsilon_{m}$ is given as
 \begin{equation}
\label{eq:drude}
\varepsilon_m(\omega)=-\left(\frac{\omega_p^2}{\omega^2+\Gamma^2}
-\varepsilon_h \right) -
\im\frac{\omega_p^2\Gamma}{\omega(\omega^2+\Gamma^2)}
 \end{equation}
where $\varepsilon_h$ denotes the high frequency permittivity,
$\omega_p$ is the plasma frequency and $\Gamma$ accounts for metal
losses in frequency units \cite{maier}.

By virtue of Eq. \eqref{eq:drude}  we can express the
$\varepsilon_\PT$ and $\varepsilon_c$ in terms of the
frequency $\omega$. Moreover, taking  Eq. \eqref{eq:singularity}
into consideration, we may obtain the SPP resonance frequency
$\omega_{sp}$ \cite{review2005,maier,huang2007}
\begin{equation}
\label{eq:wsp}
\omega_{sp}=\sqrt{\frac{\omega_p^2}{\edr+\varepsilon_h}-\Gamma^2}
\end{equation}
It can be proven that for Drude metals the $\varepsilon_\PT$ is
always smaller than $\varepsilon_c$ for frequencies lower than
$\omega_{sp}$. Thus, according to our theoretical results we
anticipate lossless SPPs propagation for $\omega<\omega_{sp}$ and
prohibited SPPs propagation for $\omega>\omega_{sp}$.

\begin{figure}[h]
\includegraphics[scale=0.42]{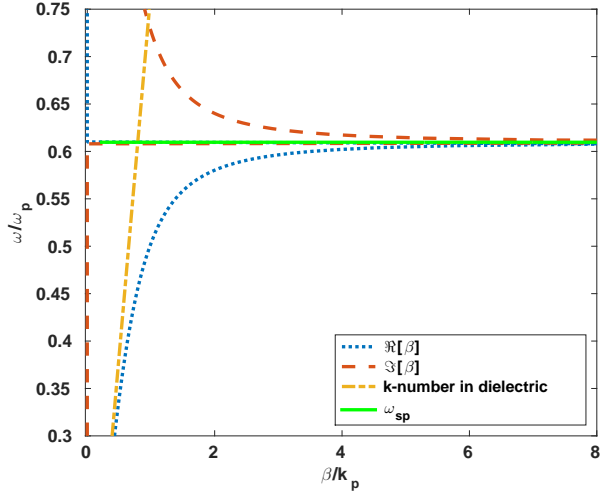}
\caption{The SPP dispersion relation $\beta$ under $\PT$ symmetry
($\edi=\varepsilon_\PT$), i.e., $\beta_0$, with respect to the
 frequency $\omega$. $\Re[\beta]$ and $\Im[\beta]$ are
indicated by dotted blue line and orange dashed line respectively.
The dash-dot yellow line is referred to the wave number of light
in the dielectric, whereas the horizontal solid green line shows
the SPP resonance frequency $\omega_{sp}$ where the interchanging
between $\Re[\beta]$ and $\Im[\beta]$ appears. $k_p=\omega_p/c$ is
used as normalized unit of wavenumbers and  $\omega_p$ as normalized unit for frequencies as well.}\label{fig:ptDispersion}
\end{figure}

In order to verify our theoretical predictions, we calculate the
SPP dispersion relation for an interface consisting of  silver
with $\varepsilon_h=1$, $\omega_p=1.367~10^{16}Hz$,
$\Gamma=1.018~10^{14} Hz$ and  silica glass with $\edr=1.69$
 {and $\edi=\varepsilon_\PT$.} The frequency values are
confined in the regime imposed by the inequality in Eq.
(\ref{eq:conditionForRealGain}). In Fig. \ref{fig:ptDispersion} we
plot the real (dotted blue line) and imaginary (orange dashed
line) part of the normalized SPP dispersion relation $\beta_0/k_p$
($k_p\equiv \omega_p/c$) with respect to the normalized frequency
$\omega/\omega_{p}$; the yellow dash-dot line shows the wave
number ($k$-number) in the dielectric, and the resonance frequency
$\omega_{sp}$ is represented by the horizontal green solid line
where the interchange between $\Re[\beta]$ and $\Im[\beta]$
appears. We observe, indeed, that for $\omega<\omega_{sp}$ the
imaginary part of $\beta$ vanishes while for $\omega>\omega_{sp}$ the SPPs wave number is purely real. Subsequently, in the
vicinity of $\omega=\omega_{sp}$ a phase transition from lossless
to prohibited SPPs propagation is expected (see Section
\ref{sec:simulations}).

\begin{figure}[h]
\includegraphics[scale=0.35]{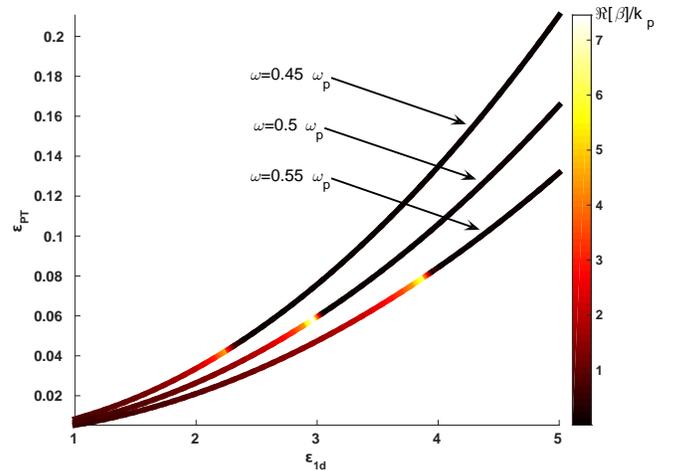}
\caption{The $\PT$ gain $\varepsilon_{\PT}$ with respect to the real part of the dielectric $\varepsilon_{1d}$ is plotted for three different frequencies $\omega$. The colors on each curve represents the magnitude of the $\Re[\beta]$ normalized to $k_p$. We observe that the more dense the dielectric is the more gain is needed  to achieve lossless SPP propagation. For all frequencies there is a critical gain for which the $\Re[\beta]$ transmits suddently from  high magnitude (bright colored line) to zero value (black line), showing gain saturation and a phase a transition from undapted SPPs to forbidden propagation.}\label{fig:edmap}
\end{figure}

Fig. \ref{fig:ptDispersion} highlights the relation between $\beta$  and the metal permittivity, and demonstrates  the $\PT$ symmetry breaking point  $\varepsilon_c$, where the $\Re[\beta]$ vanishes. 
In Fig. \ref{fig:edmap} we consider a variable $\edr$ and record the dependence on it of both the magnitude  $\Re[\beta]$ in Eq. \eqref{eq:dispersion} and $\varepsilon_{\PT}$ in Eq. \eqref{eq:PTgain} for three different frequencies, namely $\omega=\{0.45\omega_p,~0.5\omega_p,~0.55\omega_p\}$. The former is represented by color lines on the complex plane defined by $(\varepsilon_{1d},\varepsilon_{\PT})$ for $x$ and $y$ axis, respectively.  Each color line corresponds to a different frequency. Fig. \ref{fig:edmap} unveils that the more dense the dielectric is the higher value of the gain we need for having undamped SPPs propagation. In addition, the $\Re[\beta]$ vanishes very suddenly as we increase the gain, verifying that at this point the $\PT$ symmetry breaks and the SPPs propagation becomes prohibited. According to  the aforementioned figure  we can tune the magnitude of the $\Re[\beta]$ as well as $\varepsilon_c$ and $\varepsilon_{\PT}$  by choosing the appropriate dielectric.

\section{Simulations}
\label{sec:simulations}

In this Section, we verify our theoretical predictions of Sections \ref{sec:mainTheory} and \ref{sec:drude}, by solving numerically the full system of ME in the frequency domain in a two dimensional space (2D) for TM polarization electric and magnetic fields. The numerical experiments have been performed by virtue of the multi-physics commercial software COMSOL.
Precisely, we explore the SPPs propagation length $L$ with respect
to $\omega$ on the interface between two semi-infinity layers,
i.e., an active dielectric and a Drude metal, recording the
desired phase transition from lossless to prohibited SPP
propagation.
We further demonstrate the lossless SPPs propagation,  analysing
the magnetic field intensity along the surface of two known in
literature configurations, the Kretschmann-Raether and the Otto
configurations \cite{review2005,maier,leonNatPhot2012}. In our
numerical experiments the frequency $\omega$  is confined in the
range $[0.3\omega_p,0.75\omega_p]$ with the integration step
$\Delta \omega=0.01\omega_p$.

Regarding the active dielectric -- Drude metal interface described
in the previous section, we conduct the near-field excitation
technique \cite{maier,basovNanoLet2011,basovNatLet2012} to excite
SPPs on the metallic surface. For this purpose, a circular EM
source of radius $R=20nm$ has been located $100nm$ above the
metallic surface acting as a point source, since the wavelength
$\lambda$ of the EM wave in the silica glass is constrained to
$\lambda>>R$ \cite{maier}.  In addition, Perfectly Matched Layers
(PML) are used as boundary conditions.

\begin{figure}
\includegraphics[scale=0.35]{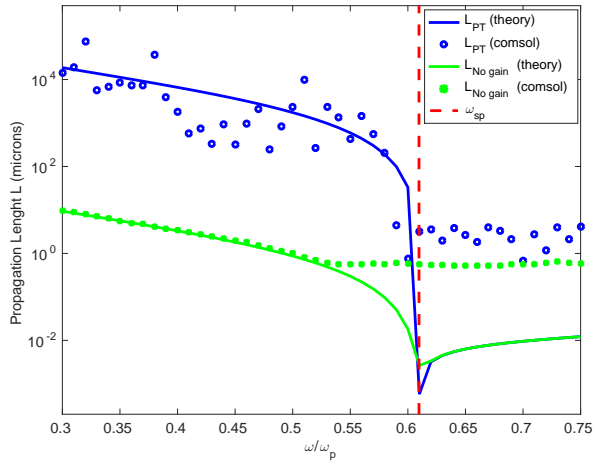}
\caption{The SPP characteristic length $L$ is demonstrated as
function of the frequency $\omega$ subject  $\PT$ symmetry (blue
line/open circles) and in the case of gainless propagation  (green
line/points). In both cases, the solid lines indicate the
theoretical prediction, whereas the points show the numerical
results obtained by COMSOL simulations. A phase transition from
lossless (large $L$) to prohibited propagation (small $L$) occurs
in the case of $\PT$ symmetry. The red vertical dashed line
indicates the resonance frequency $\omega_{sp}$, where the phase
transition takes place. The theoretical curves show deviation from
simulations for frequencies near and greater that $\omega_{sp}$,
because in this regime quasi-bound EM modes appear.} \label{fig:L}
\end{figure}

 {In Fig.\ref{fig:L} we  demonstrate, in a log-linear scale,
the propagation length $L$ with respect to $\omega$ subject $\PT$
symmetry (blue line and open circle).  For the sake of comparison,
we plot $L(\omega)$ for the gainless case (green line and filled
circles). The solid lines represent the theoretical predictions
obtained by Eq. (\ref{eq:propL}), whereas the circles indicate
COMSOL results. For the numerical calculations, the characteristic
propagation length has been estimated by the inverse of the slope
of the $log(I)$, where $I$ is the magnetic intensity along the
interface \cite{review2005,maier,mariosSPP}.
The red vertical  dashed line denotes the SPP resonance frequency
$\omega_{sp}$, in which the phase transition appears.
The graphs in Fig.  \ref{fig:L} indicate that in the presence of
the $\PT$ gain, i.e. $\varepsilon_{2m}=\varepsilon_{\PT}$, the
SPPs may travel for very long, practically infinite, distances.
Approaching the resonance frequency $\omega_{sp}$,  $L$ decreases
rapidly leading to a steep phase transition on the SPPs
propagation.
The  deviations between theoretical and numerical results in Fig.
\ref{fig:L} for frequencies near or greater than $\omega_{sp}$ are
attributed  to the fact that in the regime
$\omega_{sp}<\omega<\omega_{p}$, there are quasi-bound EM modes
\cite{maier}, where EM waves are evanescent along the
metal-dielectric interface and radiate perpendicular to this.
Consequently, the observed EM field for $\omega>\omega_{sp}$ does
not correspond to SPPs but belongs to the quasi-bound modes.}

So far,  the theoretical findings in Sections \ref{sec:mainTheory}
and \ref{sec:drude} have been successfully confirmed. We further
proceed investigating the  $\PT$ symmetry in active plasmonic
systems. We perform COMSOL simulations  based on the Total
Internal Reflection (TIR) method,  applied on the
Kretschmann-Raether and Otto configurations, separately. Within
the former,  a thin metal film is sandwiched between two
dielectrics with the incident wave hitting the denser medium. In
 Otto configuration, the denser the dielectric and the metal
sandwich a lighter dielectric.  In both configurations a type of
silica glass, with dielectric constant $\varepsilon_{D}=4$, is
used as denser passive dielectric, whereas for active dielectric (lighter medium) as well
as for metal, we use the materials described in Section \ref{sec:drude}.
 Furthermore, for the COMSOL simulations \cite{mariosSPP,altasim}, we
utilize a monochromatic plane wave source of frequency
$f=870~THz$, amounting to $40\%$ of plasma frequency $\omega_p$,
and corresponding to a wavelength $\lambda=345nm$ in the
ultraviolet regime. Again, PML are used as boundary conditions.

According to the Eq. (\ref{eq:PTgain}), the $\PT$ gain for which
SPPs propagate without losses is calculated to be
$\varepsilon_\PT=0.012$. We observe, however, that in the
numerical experiments a larger gain is needed (of the same
magnitude though), namely $\tilde\varepsilon_\PT=0.026$, for both
configurations. This  deviation from the theoretical value can be
justified if one takes into consideration that the assumption of
semi-infinitely thick metal and dielectric layers composing the
interface \cite{gainOSA2004}, under which the SPP dispersion
relation of Eq. (\ref{eq:dispersion}) holds true, is
experimentally not fully  satisfied.
\begin{figure}[h]
\centering
\includegraphics[scale=0.7]{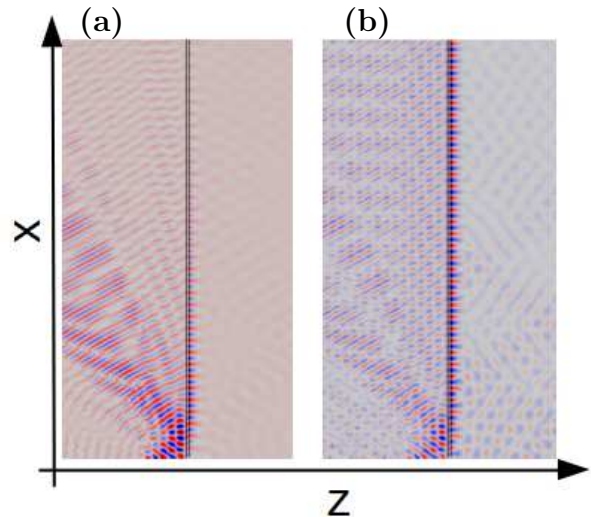}
\put(-200,186){\textbf{\large{(a)}}}
\put(-100,186){\textbf{\large{(b)}}} \caption{COMSOL simulations
show the intensity distribution of the magnetic field on a
Kretschmann-Raether configuration, for (a) $\edi=0$ (No gain) (b)
$\edi=\tilde\varepsilon_\PT=0.026$ ($\PT$
symmetry).}\label{fig:kretProp}
\end{figure}

\begin{figure}[h]
\includegraphics[scale=0.5]{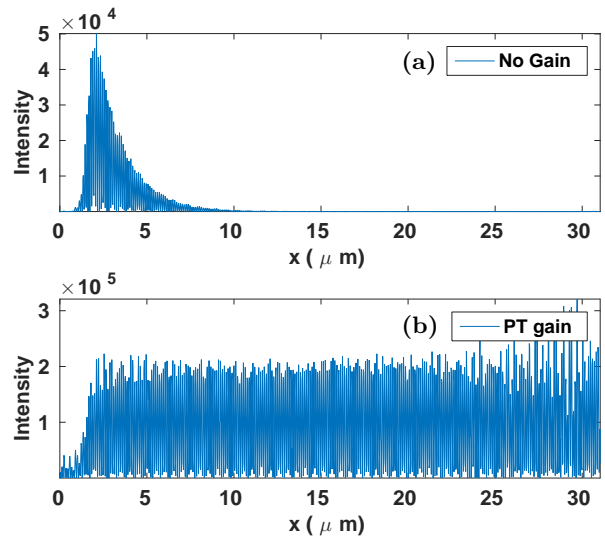}
\put(-100,182){\textbf{(a)}} \put(-100,82){\textbf{(b)}}
\caption{Characteristic profile of the magnetic field  intenstity
along the interface at Kretschmann-Raether configuration of Fig.
\ref{fig:kretProp}, for (a) $\edi=0$ (No gain) (b)
$\edi=\tilde\varepsilon_\PT=0.026$, where lossless SPP propagation
is achieved.} \label{fig:kretProfile}
\end{figure}

In the Kretschmann-Raether configuration  a thin metal of
thickness $d=45nm$ has been used for exciting SPPs. The resulting
propagation is illustrated in Fig. \ref{fig:kretProp}a under lack
of gain and in Fig. \ref{fig:kretProp}b for the $\PT$ symmetric
case. The corresponding profiles of the magnetic field intensity
along the interface are demonstrated in Fig.
\ref{fig:kretProfile}, where the lossless SPP propagation is
evident.

\begin{figure}[h]
\centering
\includegraphics[scale=0.55]{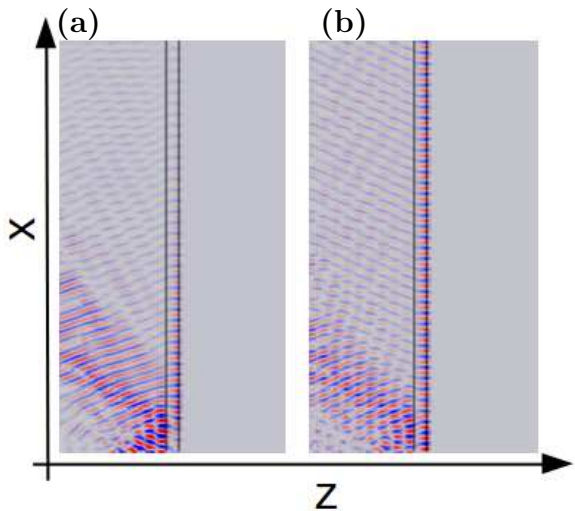}
\put(-200,184){\textbf{\large{(a)}}}
\put(-100,184){\textbf{\large{(b)}}}
\caption{COMSOL results for the intensity distribution of the
magnetic field at an Otto configuration, for (a) $\edi=0$ (No
gain) (b) $\edi=\tilde\varepsilon_\PT=0.026$ ($\PT$
symmetry).}\label{fig:OttoProp}
\end{figure}

\begin{figure}[h]
\includegraphics[scale=0.5]{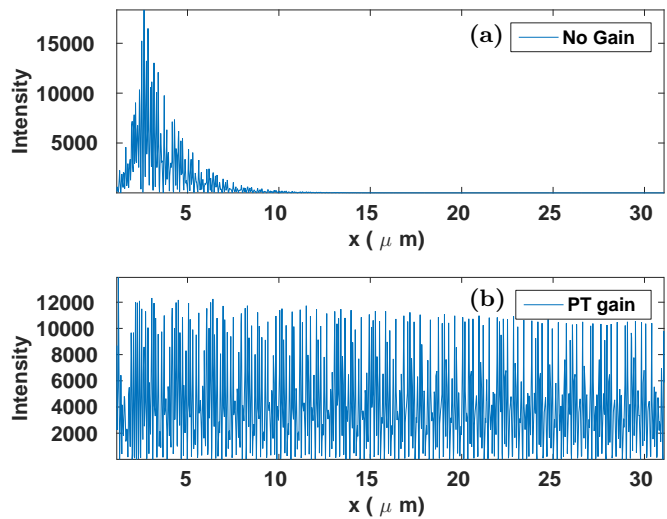}
\put(-100,185){\textbf{(a)}} \put(-100,85){\textbf{(b)}}
\caption{Characteristic profile of the magnetic field intenstity
along the interface at Otto configuration  of Fig.
\ref{fig:OttoProp}, for (a) $\edi=0$ (No gain) (b)
$\edi=\tilde\varepsilon_\PT=0.026$, where lossless SPP propagation
is achieved.} \label{fig:OttoProfile}
\end{figure}

In the Otto configuration, on the other hand, the SPPs excitation
is succeeded by means of an active dielectric of thickness $d=150
nm$ which has been used between a non-active dielectric and a
metal. By analogy to the Kretschmann-Raether configuration
experiment, we present the SPP propagation without gain in Fig.
\ref{fig:OttoProp}(a)  and for the  $\PT$ gain with
$\tilde\varepsilon_\PT=0.026$ in Fig. \ref{fig:OttoProp}(b). In
Fig. \ref{fig:OttoProfile} the corresponding profiles of the
magnetic intensity along the interface are presented, unveiling
again a clear lossless SPPs propagation in the $\PT$ case.

\section{Concluding Remarks}
\label{sec:conclusion}

Summarizing, we have investigated the role of active/gain
dielectrics in plasmonic systems. In particular, we have studied
the propagation properties of surface plasmon polaritons (SPPs)
along an interface confined by two semi-infinite layers: a
dielectric and a metal. We have calculated an exact expression
$\varepsilon_{2d}^{-}$ for the dielectric gain $\varepsilon_{2d}$,
for which the metal losses have been completely counterbalanced,
resulting to lossless SPPs propagation along the interface.
We argued that the a plasmonic system characterized by the
aforementioned lossless propagation may be related to $\PT$
symmetric systems, i.e., $\varepsilon_\PT\equiv
\varepsilon_{2d}^{-}$. Within the $\PT$ symmetry, a critical gain
$\varepsilon_c$ exists  distinguishing  between the real and imaginary part of the SPP dispersion relation. This distinction corresponds to  a phase transition from lossless to prohibited SPP
propagation.  It is remarkable that  the $\varepsilon_\PT$ as well
as the $\varepsilon_c$ depend on the optical properties of the
interface.

We applied our theory to interfaces consisting of Drude
metals and gain dielectrics demonstrating the predicted by the
theory lossless propagation  as well as the phase transition  at
the SPP resonance frequency $\omega_{sp}$. We performed numerical simulations with COMSOL software, using the near-field excitation method in order to
investigate our theory, verifying successfully all the theoretical
predictions. We also performed COMSOL simulations for two
different plasmonic configurations based on the TIR method-
Kretschmann-Raether and Otto configurations-  where  lossless SPP
propagation can be achieved.

Active metamaterials may be designed to have the desirable
frequency-dependent permittivity response, as Eq.
(\ref{eq:PTgain}) points out;  these metamaterials could be used
for the fabrication of $\PT$  symmetric plasmonic systems,
providing infinite SPPs propagation. The active metamaterials may
be used to design  $\PT$ symmetric plasmonic integrated circuits
which could transfer information in sub-wavelength scales for
large (theoretically infinite) distance, rather than a passive
plasmonic system where SPPs propagate for few micrometers.
Moreover, we demonstrated that there is a threshold in the $\PT$
gain values, above which the $\PT$ symmetry breaks and thus the system passes from lossless to prohibited propagation. The gain threshold as well as the $\PT$ gain depend
on the optical properties of the dielectric and metal,
subsequently we could control the SPPs propagation by tuning  the
dielectric constant of metal $\varepsilon_m$ or the real part of
dielectric permittivity $\edr$; for instance, the former, i.e.
$\varepsilon_m$, is usually frequency-dependent, thus we can
interchange between lossless and prohibited SPP propagation by
tuning the frequency of the incident EM wave. 

\section*{Acknowledgements}
This work was supported in part by the European Union program FP7-REGPOT-2012-2013-1 under grant agreement 316165. In addition it was partially supported  by Fondecyt grant 1120123, Programa ICM P10-030-F, the Programa de Financiamiento Basal de CONICYT (FB0824/2008), by the Ministry of Education and Science of the Republic of Kazakhstan (Contract $\#$ 339/76-2015) and the Ministry of Education and Science of the Russian Federation in the framework of Increase Competitiveness Program of  NUST $\ll$ MISiS $\gg$ (No. К2-2015-007).

\end{document}